# Vectorial near-field coupling

Martin Esmann,[1] Simon F. Becker,[1] Julia Witt,[2] Ralf Vogelgesang,[1]
Gunther Wittstock,[2] and Christoph Lienau[1]

[1] Institute of Physics and Center of Interface Science, Carl von Ossietzky University, 26111 Oldenburg, Germany

[2] Institute of Chemistry and Center of Interface Science, Carl von Ossietzky University, 26111 Oldenburg, Germany

**Abstract**

The coherent exchange of optical near fields between two neighboring dipoles[1, 2, 3] plays an essential role for the optical properties, quantum dynamics and thus for the function of many naturally occurring[4, 5] and artificial[3, 6, 7, 8, 9] nanosystems. These interactions are inherently short-ranged, extending over a few nanometers only, and depend sensitively on relative orientation, detuning and dephasing, i.e., on the vectorial properties of the coupled dipolar near fields. This makes it challenging to analyze them experimentally.

Here, we introduce plasmonic nanofocusing spectroscopy to record coherent light scattering spectra with 5-nm spatial resolution from a small dipole antenna, excited solely by evanescent fields[10, 11, 12], and coupled to plasmon resonances in a single gold nanorod[13, 14, 15]. We resolve mode couplings, resonance energy shifts and Purcell effects as a function of dipole distance and relative orientation, and show how they arise from different vectorial components of the interacting optical near-fields. Our results pave the way for using dipolar alignment to control the optical properties and function of nanoscale systems.

**Introduction**

The coherent interaction between nanoscale dipole moments, coupled via their optical near-fields over few nanometer distances, is essential for the optical properties of many types of nanostructured matter. These range from biological light harvesting complexes[4, 5, 16] to coupled semiconductor quantum dots[3, 17, 18], molecules[1, 2], plasmonic nanostructures[19, 20, 21] and combinations thereof[6, 7, 22]. Such dipole-dipole couplings account for a wide variety of apparently rather diverse physical phenomena including Förster Resonant Energy Transfer (FRET)[23], fluorescence quenching in the vicinity of metal particles[6], excitonic wavefunction delocalization or superradiance in molecular antenna systems[2, 24]. It may even proof crucial for the implementation of conditional quantum logic gates[9, 18, 25]. The active engineering of dipole-dipole interactions and the ensuing ultrafast coherent dynamics on the nanoscale thus holds promise of exciting new applications including the design of highly efficient donor-acceptor systems for organic photovoltaics[5, 26] or improved catalysts for sunlight-driven water splitting[27].

Quite generally, all these phenomena arise from the near-field coupling between two (or more) electromagnetic point-like dipoles $\mathbf{p}_1$ and $\mathbf{p}_2$, separated by a distance $R = |\mathbf{R}| = |\mathbf{r}_2 - \mathbf{r}_1|$ and driven by an external light field. Classically, their interaction energy is given as[1, 2]

$$W_{dd} = \frac{1}{4\pi\epsilon_0 R^3}\left(\mathbf{p}_1\mathbf{p}_2 - \frac{3(\mathbf{p}_1\mathbf{R})(\mathbf{p}_2\mathbf{R})}{R^2}\right), \tag{1}$$

and hence depends sensitively on dipole-dipole distance and alignment. Extracting such near-field couplings from nanostructures in their native environment at room temperature, in the presence of inhomogeneous broadening and rapid dephasing, is challenging. It requires nanometer optical resolution far below the diffraction limit to isolate the interacting dipoles and precise control over both, their distance and relative orientation.

In principle, scattering-type scanning near-field optical microscopy (SNOM) can provide the desired nanometer spatial resolution[28, 29] to record coherent light scattering spectra and sense nanoscale dipolar near-field interactions[14, 20, 30]. Earlier experiments, however, either lacked the orientational[30] or spectroscopic information[14, 20] that is necessary to unravel the interaction between the two dipole vectors.

Here, we meet these challenges by using plasmonic nanofocusing[10, 11, 31] to create an isolated, few-nanometer-sized and virtually background-free dipole emitter at the apex of a sharp gold nanotaper. We study how the light scattering from this dipole is affected by approaching it to a single gold nanorod. We directly observe the coupling-induced changes in color, linewidth and phase of the spectra as a function of tip-sample distance and relative orientation. Through analysis of Stark shifts and Purcell effects, we identify couplings induced by different vector components of the two dipoles. This advances coherent near-field spectroscopy and lays the foundation for coherent all-optical experiments with true nanometer spatial and femtosecond temporal resolution.

**Results and Discussion**

Experimentally, we create a single, spatially isolated probe dipole $\mathbf{p}_t$ by plasmonic nanofocusing as schematically shown in Figure 1a. A sharp, conical gold nanotaper is equipped with a grating coupler 50 μm above its apex of 9 nm radius. Upon focusing 300 μW of broadband laser radiation (650 nm − 900 nm) onto the coupler, surface plasmon polaritons (SPPs) are launched onto the shaft and propagate along the taper acting as a conical waveguide. The decreasing cross-section of the taper leads to a diverging in-plane mode index of the lowest order, i.e., radially symmetric surface bound SPP mode [10, 11, 32]. This is accompanied by a decrease in SPP velocity and results in the excitation of strongly localized near-fields around the apex[10, 12]. Since the mode propagates as an evanescent mode, it is converted into far field radiation only at the very apex of the taper. Here, plasmonic nanofocusing results in spatially highly localized emission with a field distribution that matches that of an isolated dipole $\mathbf{p}_t$, oriented along the taper axis[33]. This virtually background-free apex emission is collected by conventional far-field optics in a back-scattering geometry and detected with a spectrometer (see Methods for details).

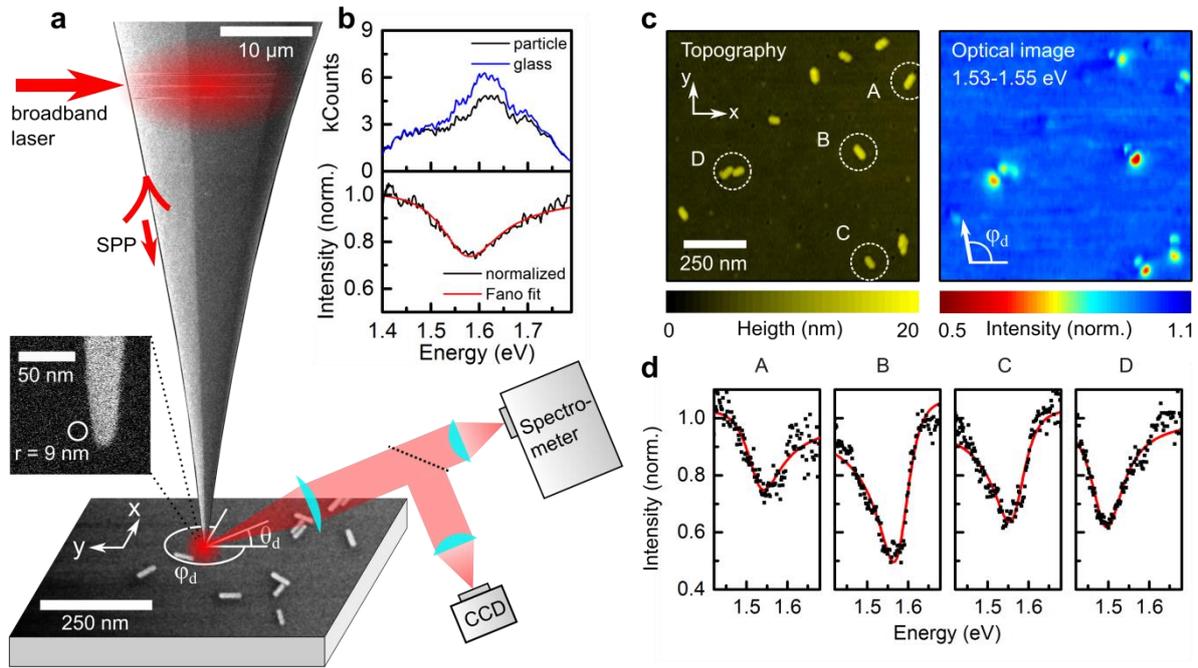

*Figure 1 Plasmonic nanofocusing scattering (PNS) spectroscopy of individual gold nanorods.* ***a,*** *Broadband SPP waves are nanofocused to the apex of a gold taper and used for local light scattering spectroscopy of $40\ nm\ x\ 10\ nm$ gold nanorods.* ***b,*** *(top) PNS spectra recorded with the nanofocusing source placed on the glass substrate (blue) and at the apex of a nanorod (black). (bottom) Normalized nanorod PNS spectrum (black) together with a fit to a Fano-type lineshape function (red).* ***c,*** *(left) Topography image of gold nanorods (Positions A-C) and a nanorod dimer (D). (right) PNS spectra were recorded at each pixel during the scan. The false color image shows the normalized scattering intensity close to the nanorod resonance ($1.54\ eV$).* ***d,*** *Normalized PNS spectra extracted at the apices of particles A-D (black) and corresponding Fano fits (red).*

As prototypical sample dipoles, $\mathbf{p}_r$, we use chemically synthesized gold nanorods of $40\ nm$ length and $10\ nm$ diameter, exhibiting a longitudinal surface plasmon (SP) resonance at $1.56\ eV$ when excited with light polarized along their long axis. The nanorods are immobilized on a glass substrate and mounted on the 3D-piezo actuator of a tuning-fork atomic-force microscope (AFM). This allows us to perform topography-tracing scans at constant tip-sample distance of $2\ nm$. At each pixel, a complete plasmonic nanofocusing scattering (PNS) spectrum between $1.35\ eV$ and $1.8\ eV$ is acquired within $20\ ms$. This way, topographic and spectral information can be directly correlated. By finely scanning the position $\mathbf{R}$ we can thus probe the coupling between the two dipoles $\mathbf{p}_t$ and $\mathbf{p}_r$, oriented orthogonally along the taper and nanorod axis, respectively. Two typical raw scattering spectra are plotted in Figure 1b (top), showing the emission from the tip apex placed either on the bare glass substrate (blue) or close to the apex of a nanorod (black). Using the substrate spectrum for normalization, we obtain the spectral response of the two coupled dipoles (bottom panel of Figure 1b

(black)). A clear reduction in scattered intensity of approximately 25% is observed around 1.6 eV as a signature of dipole-dipole coupling. We model the normalized spectrum $I(\omega)$ by a Fano-type line shape function[34], following

$$I(\omega) = \left|1 + \frac{|\mu|^2 \gamma e^{i\phi}}{\omega - \omega_0 + i\gamma}\right|^2. \qquad (2)$$

This means that we decompose the response of the system into the coherent superposition of two electromagnetic fields impinging on the detector: The broadband initial excitation emitted by the tip dipole $\mathbf{p}_t$ (to which we normalize) and secondary, coupling-induced fields, which we approximate by an effective single Lorentzian centered at an energy $\hbar\omega_0$ with a linewidth of $\hbar\gamma$ and a relative phase $\phi$ with respect to the excitation. As seen in Figure 1b (red line), this line shape describes the measured normalized spectra quite well.

A representative topography scan in a 1 μm x 1 μm sized area is shown in Figure 1c. Several nanorods (A-C) with varying in-plane orientation and a nanorod dimer (D) are readily identified. A two-dimensional map of the normalized scattered intensity, near an energy of 1.54 eV, i.e., slightly below the nanorod resonance, is shown in Figure 1d. Many, but not all rods show up as two-lobed features, displaying intensity minima at the nanorod apices. Up to 50% reduction in scattering intensity is observed for the most intense nanorods. Here, the efficient coupling to the nanorod dipole greatly reduces the amount of light that is scattered from the tip apex into the far field. In between the nanorod apex lobes, i.e. when the tip is placed over the particle center, the scattered intensity is almost equal to the substrate level. Several particles do not show up in the optical image. AFM inspection reveals geometric dimensions that are so different from the nominal values that their plasmon resonances are shifted outside the selected energy range. Generally, the scattering intensities at the two apices of the same nanorod are somewhat different. This is attributed to the asymmetric placement of the detection optics at in-plane angle $\phi_d$, indicated by the white arrow in Figure 1c (right panel).

Normalized scattering spectra of particles A-D, recorded near the rod apices, are shown in Figure 1d (symbols) together with fits to Equation (2) (red line). All spectra display a nearly absorptive lineshape, slightly red-shifted with respect to the nominal nanorod resonance. The linewidths of $\hbar\gamma \sim 50$ meV are broader than in corresponding far-field measurements. The dimer PNS spectrum is markedly red-shifted to 1.49 eV due to the hybridization of the two nanorod modes[19].

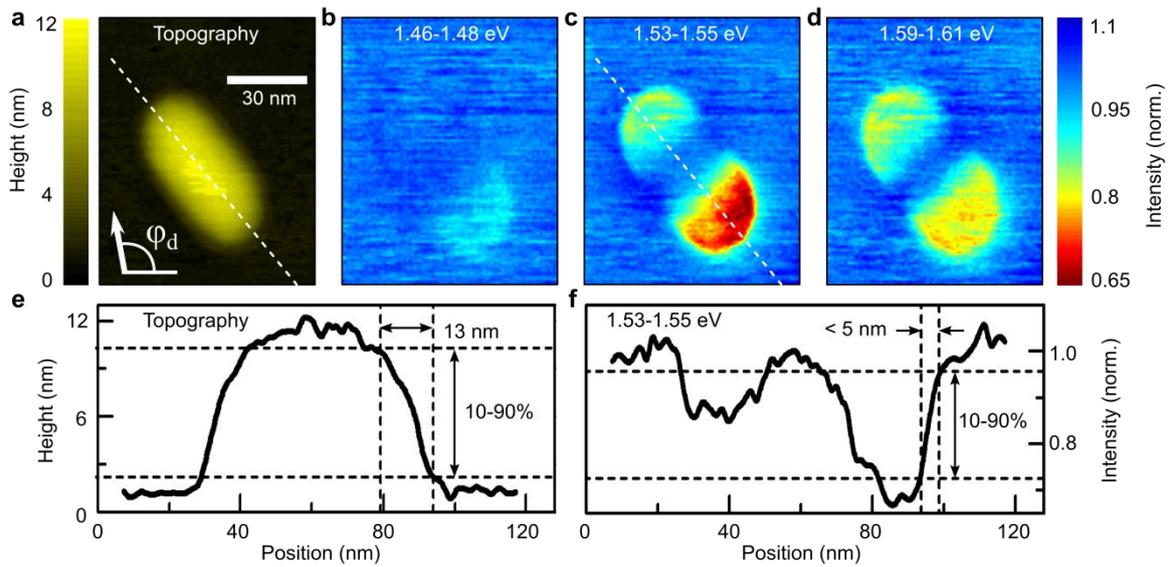

*Figure 2 High spatial resolution PNS spectra of a single nanorod. **a,** Topography of a 40 nm x 10 nm gold nanorod recorded with 1 nm x 1 nm pixel size. **b-d,** PNS spectra were recorded at each pixel during the scan. PNS intensity maps in selected energy ranges showing the optical mode profile of the nanorod were extracted from the spectra and normalized to the intensity recorded on the bare substrate. Pronounced signal reductions of up to 40% are observed at the nanorod apices. The orientation of the detector is indicated by the white arrow in panel a. **e-f,** Cross-sections extracted along the white dashed lines in panels a and c, respectively. By applying a 10-90% criterion, we estimate an optical resolution of better than 5 nm.*

To study the tip-nanorod coupling in more detail, we show high-resolution PNS images of a nanorod in Figure 2a-d. Next to the topography, three spatial maps of normalized scattered intensities, integrated over 20 meV wide spectral windows, are displayed. The images resemble the optical mode profile of the fundamental longitudinal SP resonance of the nanorod. The reduction in scattering intensity of up to 40% is more pronounced near the lower end of the nanorod facing away from the detector. A clear maximum in image contrast is seen near the nanorod resonance. Using a 10-90% criterion, we estimate an optical resolution of 5 nm from the cross sections (Figure 2e,f) along the nanorod long axis (dashed white lines in panels a and c). This is a direct measure for the strong near-field confinement at the tip apex upon plasmonic nanofocusing.

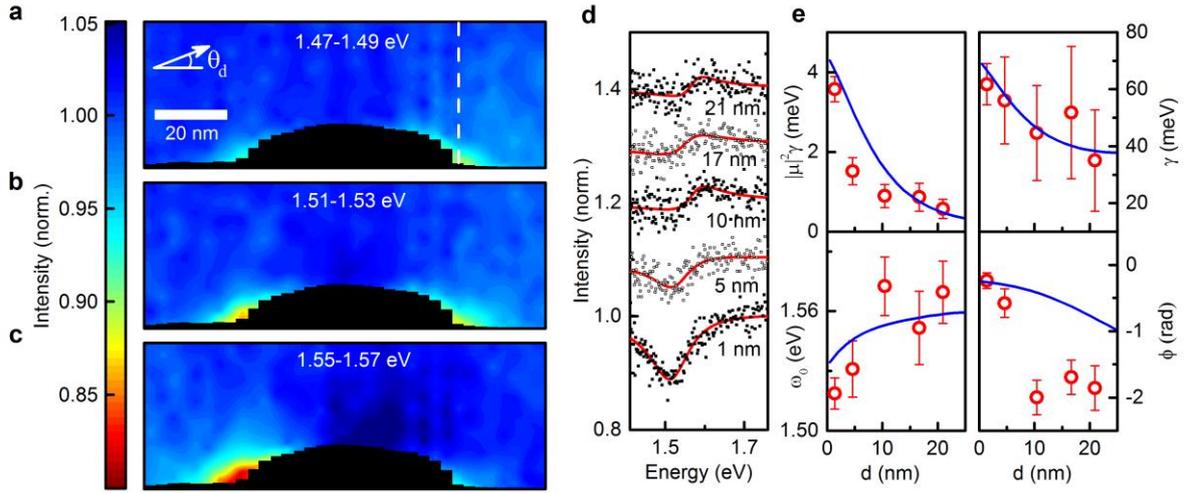

*Figure 3 Approach curve line-scan across an individual gold nanorod. **a-c,** Normalized PNS scattered intensity recorded as a function of lateral nanotaper position and tip-sample separation in three selected energy regions (topography in black). **d,** Normalized PNS scattering spectra (symbols), recorded along the white line in **(a)**, together with Fano fits (red). Subsequent curves are vertically offset by 0.1, tip-sample separations are indicated. **e,** Oscillator strength $|\mu|^2\gamma$, resonance energy $\omega_0$, linewidth $\gamma$ and Fano phase $\phi$ as a function of tip-sample distance d (red circles) extracted from the fits **(d)**. Distinct resonance shifts and Purcell effects, observed for $d < 10\ nm$, arise from the vectorial near-field coupling between tip and nanorod. The solid blue lines are coupled dipole simulations (solid blue). The variation in Fano phase from $\phi \approx 0$ to $\phi \approx -\pi/2$ accounts for the change from absorptive to dispersive lineshape.*

To study the effect of tip-sample distance on the coupling, we scanned the tip across the long axis of a single nanorod and recorded a series of distance-dependent spectra at every lateral position. Representative maps of the normalized scattering intensity are displayed in Figure 3a for three selected energy regions. The local reduction in scattering intensity is strongest when the excitation energy is tuned to the longitudinal SP resonance of the nanorod. Near-field tip-sample coupling and the resulting reduction in scattering intensity is constrained to a few-nm-sized region around the apices of the rod. The near-field contrast vanishes almost completely when removing the tip a mere 10 nm from the surface (Figure 3d). When increasing the distance by a few nm, the absorptive lineshape transforms into a dispersive shape. The results of Fano fits (Eq. (2)) to those spectra are plotted in Fig. 3e (red circles). They show a rapid increase in amplitude $|\mu|^2\gamma$ within the last 5 nm, the signature of short-range near-field dipole coupling. This coincides with a significant red shift of the scattering resonance by about $40$ meV and a concomitant increase in linewidth by about $25$ meV. The transition from dispersive to absorptive lineshape is reflected by the variation in Fano phase. These distant-dependent parameter variations are distinct signatures of multiple near-field scattering between tip and sample.

Such a lineshape analysis can also be performed for two-dimensional maps of PNS spectra, recorded at constant tip-sample distance of ∼2 nm. The resulting amplitudes $|\mu|^2\gamma(x,y)$ for a scan across a single particle are displayed in Figure 4a. Large enhancements of the scattering amplitude appear near the rod ends. The shape is different from that in Figures 2c,d and maps of optical mode profile amplitudes recorded previously with reduced spatio-spectral resolution by optical means or electron energy loss spectroscopy (EELS). To analyze the local tip-rod coupling, we first retrieve the Fano phase (Figure 4b) and note that it flips sign between the lower and upper nanorod apex. The measurements reveal precisely the local changes in resonance frequency and linewidth of the scattering resonance that are induced by the tip-nanorod coupling. Evidently, a clear Purcell effect, i.e. a coupling-induced linewidth broadening, is observed when approaching the tip to the rod apices (Figure 4c). Pronounced red shifts are seen near the outer rims (Figure 4d). They are spatially overlapping with the positions of maximum scattering amplitude as seen in the cross sections depicted in Figure 4e. Resonance shifts and Purcell effects induced by the near-field coupling between tip and nanorod are quantitatively measured for the first time.

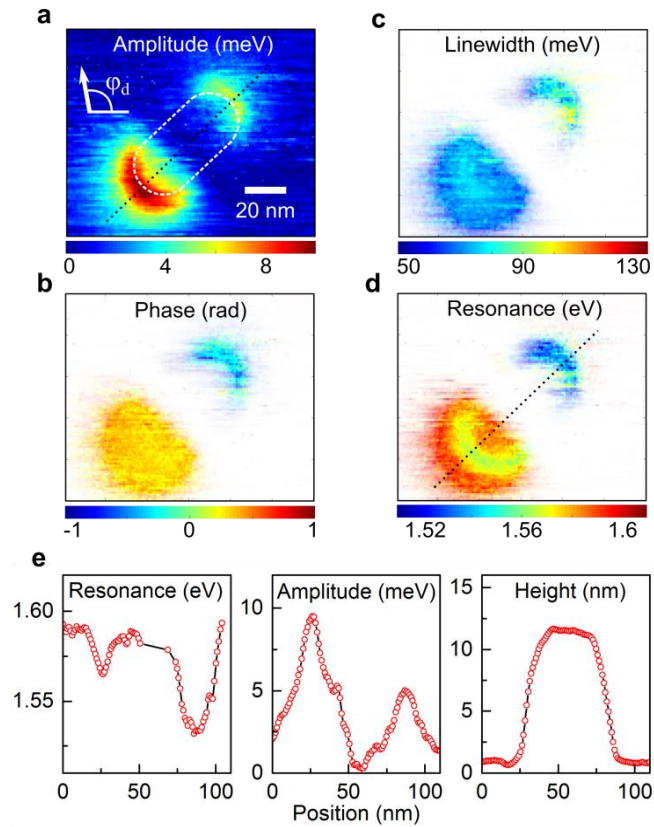

*Figure 4 Fano line analysis of nanofocusing SNOM spectra*. **a,** Map of oscillator strength $|\mu|^2\gamma$ extracted by fitting normalized scattering spectra at each pixel to Eq (2). The dashed outline shows the AFM topography. **b,** Map of fitted Fano phase $\phi$. The sign reversal represents a symmetry flip of the dispersive Fano line. **c,** Map of fitted linewidth $\gamma$ showing an increase in linewidth by ~20meV at the nanorod's apices. **d,** Map of fitted resonance energy $\omega_0$. A red-shift is observed in two crescent-moon shaped lobes at the nanorod apices. **e,** Cross-sections through $|\mu|^2\gamma$, $\omega_0$ and the topography along the dashed black line in panel a.

To account for these observations, we chose the conceptually simplest conceivable approach, representing both tip and nanorod as local dipoles, which display Lorentzian-shaped spectral resonances and are coupled via their near- and far-fields (see Methods for details). We describe the tip by a point-like polarizability tensor $\overleftrightarrow{\alpha^t}$ with a dominant longitudinal component $\alpha_{zz}^t(\omega)$ oriented along the tip axis, and weaker, off-resonant transverse components $\alpha_{xx}^t = \alpha_{yy}^t \ll \alpha_{zz}^t(\omega)$. Recent EELS measurements[35] revealed a broadband response along the tip axis which we account for by taking a resonance energy $\hbar\omega_t = 1.55$ eV and a damping of $\hbar\gamma_t = 0.33$ eV. We assume that the tip dipole is excited by both the nanofocused external laser field and the secondary field that is emitted by the nanorod.

The optical response of the nanorod is fully dominated by its longitudinal SP resonance. This can be accurately modelled by a single element $\alpha_{xx}^r(\omega)$ of its polarizability tensor $\overleftrightarrow{\alpha^r}$ with a resonance energy of $\hbar\omega_r \simeq 1.55$ eV and a linewidth of $\hbar\gamma_r \simeq 0.04$ eV[15]. To realistically model the optical mode profile of the rod, we introduce a non-local linear polarizability density that is extended along its long axis. We assume that this polarizability interacts only with the electric field that is emitted by the tip but not directly with the external laser field. The induced dipole density in the rod then generates a secondary field which acts back on the tip dipole and modifies the tip dipole radiation in the near and far field. We finally assume that the detector collects the sum of the fields emitted by tip and nanorod and, additionally, a weak background field which likely arises from the excitation of higher order angular momentum modes of the taper which are not nanofocused to the apex.[36]

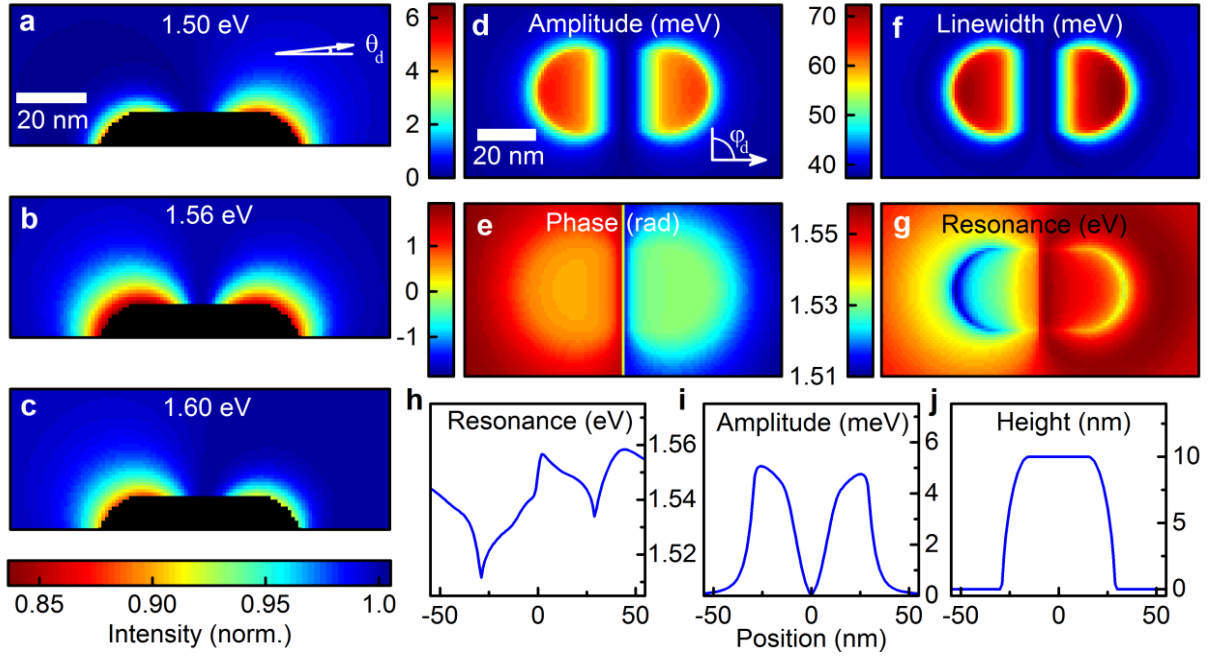

*Figure 5 Coupled dipole simulations of PNS spectra of a single gold nanorod. a-c, Normalized PNS scattered intensity recorded as a function of tip-sample distance in three selected energy regions (topography in black). The tip is modeled by a point-like polarizability tensor and the nanorod by a linear dipole density. d-g, Maps of the oscillator strength $|\mu|^2\gamma$ (d), Fano phase $\phi$ (e), linewidth $\gamma$ (f) and resonance energy $\omega_0$, (g) extracted from the coupled dipole model. h-i, Cross sections of resonance energy (h), oscillator strength (i) and sample topography (j) along the nanorod long axis. The local increase in linewidth in (f) arises from the coupling of the nanorod to the longitudinal dipolar polarization of the tip whereas strong red shifts near the rim of the rod (g,h) are induced by the nanorod coupling to the transverse tip polarizability.*

As illustrated in Figure 5a-c, this model can satisfactorily describe the approach curve scans (Figure 3a) at selected wavelengths near the nanorod resonance and it reproduces the pronounced localization of the optical near field near the two apices. As in the experiment, the maps are slightly asymmetric with respect to the rod center due to the asymmetric placement of the detector. Scattering spectra as a function of tip-sample distance $d$ (see Fig. S3 of the Supporting Information), show the same transition from dispersive to absorptive lineshapes as seen in Figure 3b. This transition can be intuitively understood when decomposing the dipole-dipole interaction between tip and sample in a perturbation series of multiple tip-sample scattering orders (denoted by a superscript). For large distances, $d >$ 20 nm, solely broad-band light scattering from the tip with a spectrum $\mathbf{E}_t^{(0)} \propto \overleftrightarrow{\mathbf{G}} \overleftrightarrow{\alpha_{zz}^t} \mathbf{E}_0$ ($\mathbf{E}_0$: nanofocused SPP field, $\overleftrightarrow{\mathbf{G}}$: Green's dyadic) is detected and no signatures of nanorod emission are found, in agreement with the experiment. This confirms our assumption that indeed only the taper is excited by plasmonic nanofocusing. It also implies that far-field coupling between tip and nanorod and hence retardation effects are negligible under our experimental conditions. For $d < 20$ nm, the near-field coupling between tip and nanorod becomes sufficiently strong to result in measurable changes of the scattering spectrum. The tip now induces the emission of a secondary dipole field $\mathbf{E}_r^{(1)} \propto \overleftrightarrow{\mathbf{G}\alpha^r}\overleftrightarrow{\mathbf{G}\alpha_{zz}^t}\mathbf{E}_0$ which interferes with the tip field on the detector. At the nanorod resonance $\omega_r$, both fields are phase-shifted by $\pm 90°$, resulting in a dispersive Fano-like interference spectrum (Figure 3d). Scattering is enhanced on the high-energy (low-energy) side of the rod resonance when positioning the tip on the rod side facing (opposing) the detector, respectively. This sign flip arises from the change in sign of the lateral ($x$-)component of the tip near field that couples to the nanorod (Supporting Informaton, Figure S2h). In this first-order perturbative limit, the back-action of the rod field on the tip is negligible. This implies that the presence of the tip does not perturb the nanorod spectrum. Hence, resonance energy and linewidth deduced from the Fano-analysis are equal to those of a conventional far-field measurement.

This changes as higher order tip-sample interactions become important. As a second order perturbation, the nanorod field that couples back to the tip induces a dipole oscillation along the taper axis which yields the additional field $\mathbf{E}_t^{(2)} \propto \overleftrightarrow{\mathbf{G}\alpha_{zz}^t}\overleftrightarrow{\mathbf{G}\alpha^r}\overleftrightarrow{\mathbf{G}\alpha_{zz}^t}\mathbf{E}_0$. Since the tip resonance is close to $\omega_r$, this field is phase-shifted by another $\pm 90°$ with respect to $\mathbf{E}_r^{(1)}$ and thus by $180°$ compared to $\mathbf{E}_t^{(0)}$. This destructive interference leads to a scattering spectrum with an absorptive line shape if the amplitude of $\mathbf{E}_t^{(2)}$ exceeds that of the field emitted from the rod. The relative amplitude of both fields strongly depends on the detector position. In our experiments, the detector is put at a small angle of 20° to the sample plane. This ensures efficient collection of tip scattering while suppressing rod emission. Hence, for small tip-sample distances the amplitude of $\mathbf{E}_t^{(2)}$ indeed exceeds that of $\mathbf{E}_r^{(1)}$. This explains the transition from dispersive to absorptive line shape (Figure 3a). However, no line broadening or spectral shifts occur at this level of approximation. Their observation inevitably requires higher order scattering processes and - thus - multiple scatterings between tip and sample. When restricting the tip polarizability to the dominant longitudinal component $\alpha_{zz}^t$, the narrow rod SP resonance spectrally overlaps the broad tip resonance. Hence, the coupling leads to a broadening of the scattering spectra, i.e. a Purcell effect. The tip increases the local optical density of states at the position of the nanorod and therefore the radiative damping of the nanorod. This can account for the increase in linewidth $\gamma$ of the scattering resonance that is seen when reducing $d$. To account for the substantial line broadening that is seen experimentally, high orders ($> 10$) of perturbation theory are needed.

The quasi-scalar **transverse coupling** discussed so far originates from a single component, $G_{xz}$, of the Green's tensor. It can account for the coupling-induced radiative broadening of the PNS spectrum, yet fails in reproducing the spectral line shifts at the rim of the nanorods (Figure 4d). To account for these shifts, a coupling of the nanorod to a blue-shifted resonance is needed that has little overlap with the rod spectrum. In this case, dipole coupling results in an optical Stark shift, red-shifting the rod resonance when coupling to a blue-detuned tip mode. Such Stark shifts are evidently seen near the outer rim of the rod in Figure 4d with a spatial dependence that is very different from that of the total scattering amplitude. A blue-shifted tip resonance arises from the transverse polarizability $\alpha_{xx/yy}^t$ [37] to which the rod dipole can couple via the component $G_{xx}$ of the Green's tensor. This **longitudinal coupling** induces lateral tip dipole oscillations which can either emit light into the far field or couple back to the rod dipole. It requires mode overlap between the $x$-components of the nanorod and the tip field and thus becomes maximum near the nanorod ends. When including the transverse polarizability and its coupling to the rod dipole in the simulations, we see pronounced red shifts of about 30 meV near the two outer rims of the nanorod (Figure 5g) agreeing well with the measurements in Figure 4d. Importantly, this longitudinal coupling induces red shifts on both sides of the rod. Hence, we conclude that the Purcell effects and spectral red shifts seen in Figure 4 reflect different aspects of **vectorial** near-field dipole coupling. While Purcell effects are induced by the coupling between the nanorod and the dominant longitudinal tip polarizability, strong red shifts of the scattering reflect the coupling to the weaker, spectrally off-resonant, transverse tip polarizability. The spectral shifts that are observed experimentally are of the same order as the width of the nanorod SP resonance but do not exceed it. Hence, all results are within the intermediate coupling regime and may thus be described with the perturbative Born series expansion of the scattering. Yet, the coupling is sufficiently strong to require up to several tens of scattering orders to quantitatively model the observed spectra. Hence, the Born series expansion, commonly employed for the interpretation of near-field images[14, 20, 37], becomes impractical and a self-consistent solution of the scattering problem is required.

In summary, we have used plasmonic nanofocusing to create an isolated, spectrally broadband and essentially background-free light source with 5-nm dimensions at the very apex of a conical gold taper. We have used this light source to study coherent light scattering from the junction between a tip and a single gold nanorod with few-nanometer spatial resolution. The recorded scattering spectra reveal unique signatures of vectorial near-field coupling between tip and nanorod dipoles, in particular Purcell effects resulting from the dominant longitudinal component of the tip polarizability and pronounced Stark shifts near the rim of the nanorod that are assigned to the coupling to its off-resonant transverse component. Such near-field couplings are ubiquitous in hybrid nanoscale systems play an important role for their function, for instance in plasmonics, photocatalysis or artificial photosynthesis. Hence, vectorial coupling phenomena, studied here for a plasmonic model system, are of fundamental relevance for the transport of energy in nanoscale systems. Plasmonic nanofocusing offers a new platform for sensing and manipulating these couplings with high precision paving the way for a coherent control of dipole-dipole coupling on the nanoscale.

**Methods**

Nanorod sample preparation: BK7 glass substrates were cleaned with ethanol and then by submersion in Piranha solution (3 volume parts 98% $H_2SO_4$ and 2 volume parts 30% $H_2O_2$) for 40 min, thorough rinsing with deionized water and blow drying. The substrates were then hydroxilated by $UV/O_3$ cleaning. Chemically synthesized and polymer-coated gold nanorods were purchased from NanoPartz Inc., Loveland CO, USA (suspension with specified concentration of $1.45 \times 10^{14}$ nanorods per mL). 2 µL of the suspension were diluted with 10 mL of dried ethanol. An aliquot of 1 mL of this suspension was further diluted with 9 mL of dried ethanol. The pretreated glass slides were then placed in a PTFE container with the second nanorod dispersion for 3 h. After retrieval, the substrates were rinsed with deionized water and blow dried in an argon stream.

Nanofocusing SNOM taper preparation: Gold nanotapers were produced using the method described in Ref. [12]. Gold wires of 125 µm diameter (Advent Research Materials) were annealed at 800°C for eight hours under argon atmosphere followed by electrochemical AC-etching (3 kHz rectangular voltage, 7.5 V peak-to-peak, 250 mV DC offset, 90% duty cycle, platinum counter-electrode) in concentrated hydrochloric acid (37%). Etched tapers were rinsed in ethanol. Three-line grating couplers (1.26 µm period, 100 nm width, 200 nm depth) were produced by Focused $Ga^+$ ion beam lithography (FEI, Helios Nanolab 600i) 50 µm above the taper apex of 10 nm radius (determined by scanning electron beam microscopy). For optimizing the optical coupling bandwidth, the grating lines were slightly inclined with respect to each other[38] resulting in a 200 nm change in grating period across the taper. The acceptance bandwidth of the grating coupler limits the spectral bandwidth of the raw PNS spectra in Figure 1b.

Plasmonic nanofocusing scattering (PNS) spectroscopy: SPPs are launched onto the shaft of a nanofocusing SNOM taper by illuminating a grating coupler with a broadband supercontinuum laser (Fianium WL-SC-400-4). This results in a spatially isolated and spectrally broadband nano light source at the 10 nm sized taper apex. A halfwave plate is used to adjust the polarization state of the light impinging onto the grating coupler to p-polarization for highest SPP coupling efficiency. For launching SPPs we use a microscope objective with a numerical aperture of $NA = 0.2$, which is inclined by 13° out of the sample plane avoiding direct illumination of the sample (cf. Figure 1a). The back-scattered optical signals are detected in reflection geometry by a separate collection objective at an angle of $\theta_d \sim 20°$ out of the sample plane with a numerical aperture of $NA = 0.35$. After collection, the scattered light is spectrally dispersed in a monochromator (Princeton Instruments, Acton SP2500) and spectra are recorded with a liquid nitrogen-cooled CCD camera (Roper Scientific, Spec-10:100BR). A second camera collects a real image of the emission from the taper apex to ensure that the light source is spatially isolated and no direct scattering off the grating coupler is collected.

Tip-sample distance regulation is realized with a custom-build tuning fork based AFM similar to the setup described in Ref. [39]. The nanofocusing SNOM taper is glued to one prong of a quartz tuning fork which is electrically driven in resonance (TTI, TG1010A) to oscillate normal to the sample surface with a peak-to-peak amplitude of ~1 nm. The resulting piezoelectric current is converted by a custom-built transimpedance amplifier and demodulated in a lock-in amplifier (FEMTO, LIA-MVD-200H). A custom-built proportional integral controller adjusts the position of the sample on a 3D piezo actuator (PI, P-363.3CD) keeping the quadrature signal $X$ of the lock-in constant and thus maintains a tip-sample distance of ~2 nm. The other two axes of the 3D piezo actuator are used to laterally raster scan the sample surface.

Figure 1 shows data obtained from a scan of a 1 μm x 1 μm size area of the sample performed at a constant tip-sample distance of 2 nm. At each 10 nm x 10 nm pixel of the scan, a full PNS spectrum was acquired within an integration time of 20 ms. The data in Figure 2 result from a high-resolution scan with a pixel size of 1 nm x 1 nm recorded with the same integration time. The distant-dependent PNS spectra in Figure 3 have been recorded by approaching the tip at constant speed towards the surface while integrating the scattered signal for 20 ms. Each approach is stopped when reaching a tip-sample distance of 2 nm, retracting the tip and moving to an adjacent lateral position. No further drift correction was applied.

Data treatment: The spectra shown in Figure 1 and Figure 2 were obtained by normalizing each PNS spectrum to a spatially averaged reference PNS spectrum recorded by placing the nanotip on the bare glass substrate. No further background suppression has been applied. Slow spatial variations in reference PNS spectrum over large sample areas were accounted for by fitting a low-order polynomial to the background before normalization. A slightly more complex normalization procedure was applied to the PNS spectra in Figure 3 and 4 where the background scattering from the glass substrate is dependent on the tip-sample distance. Distance-dependent background PNS spectra have first been recorded on the left and right side of the nanorod. Distance-dependent background spectra within the nanorod region are then obtained by linear interpolation along the lateral axis and used for calculating the local normal PNS spectra. The PNS intensity maps in Figures 1-3 are obtained by averaging the PNS spectra over the 20 meV wide region depicted in the respective figure. Fits of the PNS spectra to Equation (2) were obtained by nonlinear least-squares fitting.

Coupled dipole simulations: A coupled-dipole model has been used to analyze the experimental data. In this model, both tip and nanorod are represented as local dipoles with Lorentzian-shaped spectral resonances. We assume that the tip response can be described by a point-like polarizability tensor $\overleftrightarrow{\alpha^t}$ with a dominant longitudinal component $\alpha_{zz}^t(\omega) = |\mu_z^t|^2 L(\omega, \omega_t, \gamma_t)$. The lineshape function is $L(\omega, \omega_t, \gamma_t) = -\frac{1}{\hbar}\left(\frac{1}{\omega - \omega_t + i\gamma_t} - \frac{1}{\omega + \omega_t + i\gamma_t}\right)$. We also allow for much weaker transverse components

$\alpha_{xx}^t = \alpha_{yy}^t \ll \alpha_{zz}^t(\omega_t)$. We denote with $\overleftrightarrow{\alpha_{zz}^t}$ the tensor neglecting these off-resonant transverse components. The frequency dependence of the longitudinal polarizability is modeled by taking a resonance energy of $\hbar\omega_t = 1.55$ eV and a damping of $\hbar\gamma_t = 0.33$ eV. The linewidth is chosen in agreement with recent EELS measurements on conical tapers indicating a very broadband spectral response of the field localized near the apex. The effective dipole moment $\mu_z^t$ gives a maximum field enhancement factor of ~7 at the very apex of the taper which matches previous experiments quite well[40].

We simulate the transverse tip polarizability as that of a small spherical gold particle with 10 nm diameter exhibiting a SP resonance near 550 nm. Its field enhancement at the frequency of the nanorod resonance is six times smaller than that for the longitudinal polarizability component. This effective point-dipole is located at position $\mathbf{r}_t$, 3.3 nm above the center of a 10 nm-radius sphere that represents the taper apex. Both the external field $\mathbf{E}_0(\omega, \mathbf{r}_t)$ which is nanofocused to the apex and the field $\mathbf{E}_r(\omega, \mathbf{r}_t)$ emitted by the nanorod induce coherent dipole oscillations $\mathbf{p}_t(\omega, \mathbf{r}_t) = \overleftrightarrow{\alpha^t}(\omega)\big(\mathbf{E}_0(\omega, \mathbf{r}_t) + \mathbf{E}_r(\omega, \mathbf{r}_t)\big)$ in the tip. These dipole oscillations create electromagnetic near- and far-fields $\mathbf{E}_t(\omega, \mathbf{r}) = \frac{\omega^2}{\epsilon_0 c^2} \overleftrightarrow{\mathbf{G}}(\mathbf{r}, \mathbf{r}_t, \omega)\mathbf{p}_t(\omega, \mathbf{r}_t)$ at position $\mathbf{r}$. For simplicity, we use the Green's dyadic $\overleftrightarrow{\mathbf{G}}$ of free space, neglecting the influence of the glass substrate.

The response of the longitudinal dipolar nanorod resonance is represented by a single element $\alpha_{xx}^r(\omega)$ of the polarizability tensor $\overleftrightarrow{\alpha^r}$. Its spectral lineshape is modeled by a Lorentzian with a resonance energy of $\hbar\omega_r \simeq 1.55$ eV and a linewidth of $\hbar\gamma_r \simeq 0.04$ eV[15]. These parameters account well for far-field spectra of single nanorods recorded in our laboratory. To realistically describe the optical mode profile of the nanorod, we consider a non-local linear polarizability density $\alpha_{xx}^r(x, x', \omega) = |\mu_x^r|^2 L(\omega, \omega_r, \gamma_r)\rho(x, x')$ with $\rho(x, x') = \frac{1}{N}\cos(\pi x/L')\cos(\pi x'/L')$, $N = \frac{\pi}{2L'}\frac{1}{\sin(\pi L/L')}$ and $L' = 2L + 4R$[41]. This effective polarizability is positioned in the center of a cylinder of length $2L = 30$ nm and of radius $R = 5$ nm capped with two hemispheres of radius $R$ and centered at the origin. We assume that this nanorod polarizability interacts only with the electric field that is emitted by the tip inducing a linear dipole density $\mathbf{p}_r(\omega, x) = \int_{-L}^{L} \overleftrightarrow{\alpha^r}(\omega, x, x')\mathbf{E}_t(\omega, \mathbf{r}_{x'}) dx'$ with $\mathbf{r}_{x'} = (x', 0, 0)$. This linear dipole density then generates a secondary field $\mathbf{E}_r(\omega, \mathbf{r}) = \frac{\omega^2}{\epsilon_0 c^2}\int_{-L}^{L} \overleftrightarrow{\mathbf{G}}(\mathbf{r}, r_{x'}, \omega)\mathbf{p}_r(x', \omega) dx'$. This field acts back on the tip dipole and modifies the tip dipole radiation in the near and far field. The detector placed at position $\mathbf{r}_d$ collects the sum of the fields $\mathbf{E}_t(\omega, \mathbf{r}_d)$ and $\mathbf{E}_r(\omega, \mathbf{r}_d)$. To quantitatively match the recorded PNS spectra, we also allow for a weak background field $\mathbf{E}_b(\omega, \mathbf{r}_d) = \frac{\omega\, b}{\epsilon_0 c^2}\overleftrightarrow{\mathbf{G}}(\mathbf{r}, \mathbf{r}_t)\overleftrightarrow{\alpha^t}(\omega)\mathbf{E}_0(\omega, \mathbf{r}_t)$ to reach the detector. We assume that this background is created by SPP scattering from unavoidable roughness at the tip shaft or due to higher order angular momentum modes of the conical taper which are not fully guided to the apex[36]. We empirically find that a background amplitude $|b|\sim4.5$ yields quantitative agreement between experiment and simulation. The intensity measured on the detector is then $I(\omega,\mathbf{r}_d) \propto |\mathbf{E}_t(\omega,\mathbf{r}_d) + \mathbf{E}_r(\omega,\mathbf{r}_d) + \mathbf{E}_b(\omega,\mathbf{r}_d)|^2$. Similar to the experiment, the detector is placed at a small angle $\theta_d \simeq 10°$ out of the sample plane. In all simulations, the in-plane angle is set to $\phi_d = 0°$, i.e., the detector is oriented along the long axis of the nanorod.

Effectively, we thus model tip and nanorod sample by two effective dipoles which are coherently coupled via their optical near and far fields. The coupled dipole equations are solved both self-consistently and in a perturbative, Born-series expansion of the scattering problem[37]. We find that the Born series expansion, commonly employed for the interpretation of near-field images[14, 20, 37], generally converges under our conditions. However, more than 10 expansion orders are needed to reproduce the Stark shifts and Purcell effects seen for tip-sample distances of less than 5 nm.

**Acknowledgements**

Financial support by the Deutsche Forschungsgemeinschaft (SPP1839 "Tailored Disorder", grants LI 580/12 and SPP1840 "QUTIF" grant LI 580/13), the Korea Foundation for International Cooperation of Science and Technology (Global Research Laboratory project, K20815000003) and the German-Israeli Foundation (GIF grant no. 1256 and 1074-49.10/2009) is gratefully acknowledged. ME thanks the Studienstiftung des deutschen Volkes (German Scholarship Foundation) for a PhD scholarship. The authors thank Vladimir Smirnov for performing supporting Finite Element Method calculations and Heiko Kollmann for providing high resolution SEM images of individual gold nanorods.

**Additional information**

Supplementary information is available in the online version of the paper. Reprints and permissions information is available online at www.nature.com/reprints. Correspondence and requests for materials should be addressed to CL (email: christoph.lienau@uni-oldenburg.de).

**Author contributions**

JW and GW prepared the nanorod samples. ME and SFB built the SNOM setup, prepared the nanofocusing SNOM tapers and conducted the SNOM experiments. ME, SFB and CL analyzed the data. ME, CL and RV performed the theoretical modelling. CL initiated the project, ME and CL prepared the manuscript. All authors contributed to the interpretation of the data and the final version of the manuscript.

**Competing financial interests**

The authors declare no competing financial interests.

# Supplementary information for "Vectorial near-field coupling"

Martin Esmann,[1] Simon F. Becker,[1] Julia Witt,[2] Ralf Vogelgesang,[1] Gunther Wittstock,[2] and Christoph Lienau[1]

[1] Institute of Physics and Center of Interface Science, Carl von Ossietzky University, 26111 Oldenburg, Germany

[2] Institute of Chemistry and Center of Interface Science, Carl von Ossietzky University, 26111 Oldenburg, Germany

1. **Electron microscopy characterization of gold nanostructures**

Plasmonic nanofocusing tapers and gold nanorods used in our study were both carefully pre-characterized by Scanning Electron Beam Microscopy (SEM). Figure S1 shows typical images of an electrochemically etched taper equipped with a three-line grating coupler (a-c) and an individual, chemically synthesized gold nanorod (d). Tapers are produced by electrochemical AC-etching following the method described in Ref. 1 (see methods of our manuscript). For efficient, scattering-free SPP guiding and nanolocalization of light at the taper apex we verify that the tapers have a smooth, defect-free surface, an opening angle of $\sim 15° - 30°$ and an apex radius of curvature below 10 nm. We ensure that the tapered region of $\sim 50$ μm length consists of a single monocrystalline grain to avoid SPP scattering into the optical far-field at grain boundaries. For the taper presented in Figure S1a-c all these requirements are met. Chemically synthesized gold nanorods were also inspected by high-resolution SEM imaging. Figure S1d shows a typical particle deposited on a conducting ITO surface to avoid SEM imaging errors due to charging effects. The particle shape is well described by a cylindrical central part terminated with two hemispherical end caps.

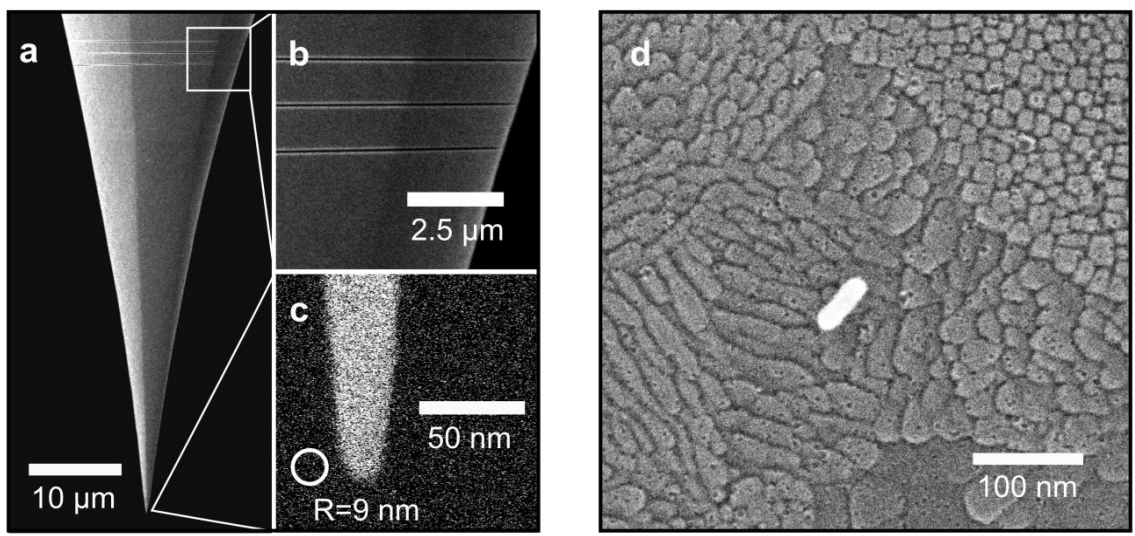

*Figure S1 Scanning electron microscopy (SEM) characterization of nanofocusing SNOM tapers and gold nanorods. a, Side-view of a monocrystalline nanofocusing gold taper produced by electrochemical etching. A grating coupler of three lines is*

milled into the taper surface 50 μm above the apex by Ga⁺ focused ion-beam lithography. **b,** Close-up view of the grating coupler with a period of 1.26 μm, optimized for SPP excitation around 1.55 eV. Grating lines are 100 nm wide, 200 nm deep and inclined with respect to each other such that the grating period varies over 200 nm across the full taper. This increases the coupling bandwidth. **c,** Close-up view of the taper apex with a radius of 9 nm, as indicated by the white circle. **d,** SEM image of a typical, chemically synthesized gold nanorod of 40 nm length and 5 nm radius, as used in our study. For SEM characterization particles were deposited on conducting ITO surfaces instead of glass.

## 2. Theoretical modeling of optical mode profiles

We quantitatively model Plasmon Nanofocusing Scattering (PNS) spectra by coupled dipole simulations. For this, we represent the spectral response of both tip and nanorod by effective Lorentzian-shaped resonances (see Methods section of our manuscript) and construct the associated spatial mode profiles from distributions of point polarizabilities. The practical implementation of these polarizabilities and the resulting optical mode profiles are shown in Figure S2.

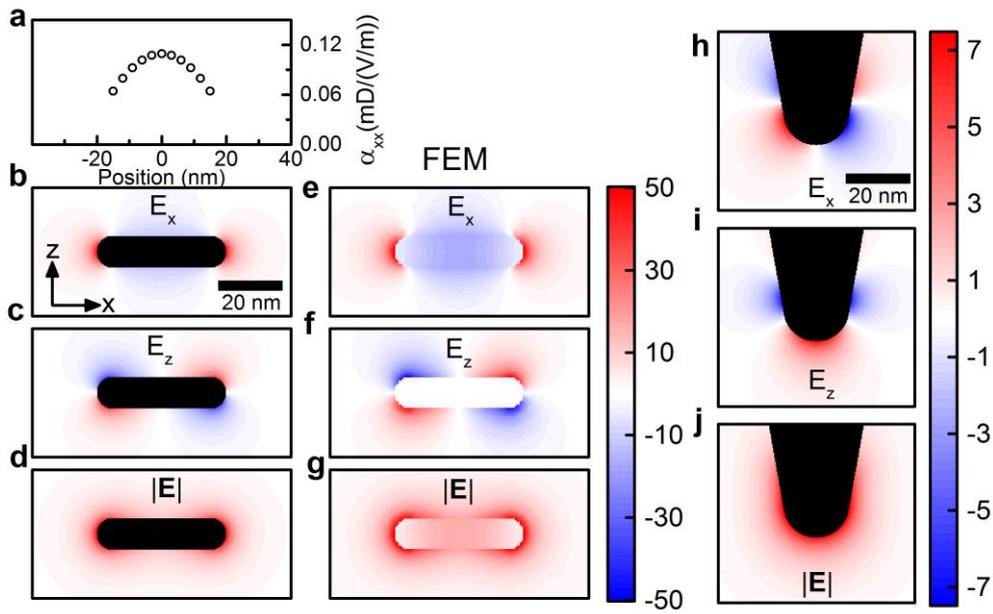

***Figure S2 Optical modes of nanorod and tip modeled by effective dipoles. a,*** *Discretized linear polarizability density* $\alpha_{xx}^r(x, \omega_r) \propto \cos\left(\frac{\pi x}{L'}\right)$ *with* $L' = 50$ *nm used to model the optical mode of a 10 nm x 40 nm gold nanorod in coupled dipole simulations.* ***b-d,*** *Resonant near-fields generated by exciting the polarizability in (a) with a plane wave propagating along the z-direction and polarized along x with an elecric field amplitude of* $1\, V/m$ *at* $\hbar\omega_r = 1.55\, eV$. *The black areas indicate the nanorod. The linear polarizability is distributed along its center axis.* ***e-g,*** *Field distribution around a gold nanorod simulated by Finite Elements Methods (FEM) under equivalent excitation conditions.* ***h-j,*** *Optical mode of a nanofocusing taper apex at* $\omega_r$ *modeled by a single polarizable point dipole located 3.3 nm above the center of the hemispherical apex cap. Near-fields are excited by an incident field polarized along z with a field strength of* $1\, V/m$. *The polarizability is adjusted to yield an experimentally realistic field enhancement of* $f_t \approx 7$ *at the apex.*

The longitudinal surface plasmon (SP) resonance of the nanorod is associated to a single component $\alpha_{xx}^r(\omega)$ of its anisotropic polarizability tensor $\overleftrightarrow{\alpha^r}$ ). To realistically model the elongated optical mode profile of the nanorod, we introduce a nonlocal linear polarizability density that is extended along the nanorod's $x$-axis and has the shape $\alpha_{xx}^r(x, x', \omega) = |\mu_x^r|^2 L(\omega, \omega_r, \gamma_r) \rho(x, x')$ with $\rho(x, x') = \frac{1}{N} cos(\pi x/L') cos(\pi x'/L')$, $N = \frac{\pi}{2L'} \frac{1}{sin(\pi L/L')}$ and $L' = 2L + 4R$ ($2L = 30$ nm and $R = 5$ nm)² and $L(\omega, \omega_r, \gamma_r)$ the Lorentzian line-shape as defined in our manuscript. For the practical implementation of this polarizability in a coupled dipole simulation, we use a discretized, linear chain of 11 polarizable points at equally spaced positions along the cylindrical part of the nanorod, as shown in Figure S2a. Here, the polarizability is shown at the maximum of the spectral resonance ($\hbar\omega_r = 1.55$ eV). The fields around this dipole chain resulting from a homogeneous excitation field of 1 V/m polarized along $x$ are shown in (b-d). The dipole chain describes the optical field of the antenna mode everywhere outside the nanorod of 10 nm diameter and 40 nm length, which is indicated by the black area. Close to the nanorod apices the optical field of the dipole chain exhibits strong components both in the $x$ and $z$ direction, which couple to the polarizability of the tip apex. We independently verify our nanorod model by Finite Element Method (FEM) simulations. The resulting fields under equal excitation conditions as for the dipole chain are plotted in Figure S2e-g. The resulting field distribution of the antenna mode outside the nanorod is virtually the same as the one resulting from the linear dipole chain (b-d). The fields in (e-g) are again shown at the nanorod resonance of $\hbar\omega_r = 1.55$ eV. We observe a field enhancement of $f_r \approx 50$ at the nanorod apices. This value was used to normalize the overall resonant polarizability of the nanorod shown in panel a, resulting in a resonant polarizability of $\alpha_{xx}^r(\omega_r) \approx 1 \frac{mD}{V/m}$. Similarly, we used FEM simulations to independently verify the value for the nanorod resonance linewidth of $\hbar\gamma_r \approx 0.04$ eV used in our coupled dipole simulations.

The spatial mode properties of the taper apex are well captured by a single polarizable point dipole, placed 3.3 nm above the center of the hemispherical apex with an anisotropic polarizability tensor $\overleftrightarrow{\alpha^t}$ ($\alpha_{zz}^t \gg \alpha_{xx}^t = \alpha_{yy}^t$). The resulting resonant field distribution around the tip apex (excitation along z, at 1.55 eV with incident field amplitude 1 V/m) is shown in panels (h-j). Since the excitation exclusively couples to the polarizability component $\alpha_{zz}^t(\omega)$, the fields here reflect the optical mode of the tip upon initial excitation by the nanofocused SPP field. Our experiments indicate that SPP nanofocusing results, in addition to the localized field at the taper apex, in a spatially less localized background field with a spectral characteristic matching that of the incident field. The background field likely arises from the excitation of higher order angular momentum modes of the taper[3] and unavoidable tip imperfections.

As for the antenna mode of the nanorod, the resulting point dipole describes the fields generated everywhere outside the tip (black area) reasonably well. We normalize the tip polarizability such that an experimentally realistic[4] value of $f_t \approx 7$ for the field enhancement at the very apex is obtained. The coupling of the tip field to the nanorod is mediated by the $x$-component of the dipole field shown in panel h.

## 3. Coupled-dipole modeling of near-fields and PNS spectra

With the discretize dipole distributions generating the optical modes of tip and sample at hand, we set up a coupled dipole model simulation. This allows us to model both the coupled near- and far-fields of the tip-nanorod system and hence to simulate full PNS spectra and maps of scattered intensities. Furthermore, by least-squares fitting to the Fano-line defined in Equation (2), we obtain simulated Fano parameter plots such as those shown in Figures 3e and 5d-g of the main text. In Figure S3 we show simulated coupled near-fields and normalized PNS spectra of a single tip-sample approach curve with the lateral tip position close to the right nanorod apex. The multiple scattering problem was solved self-consistently.

In panel (a) we show the resulting near-field amplitudes at the unperturbed resonance energy of the nanorod, $\hbar\omega_r = 1.55$ eV, for three different tip-sample distances indicated at the upper left of each panel. An additional minimum distance of 3 nm was included in these values to account for the minimum tip-sample distance accessible by AFM regulation in our experiments. All model parameters are chosen as described in the corresponding Methods section of our manuscript. In the model only the tip apex is excited by an external field $\mathbf{E}_0$ pointing along the tip axis with an amplitude of 1 V/m. Correspondingly, for large tip-sample distances around 30 nm mainly the tip mode is excited by the incident excitation, resulting in a field enhancement of $f_t \approx 7$. The nanorod mode is only weakly present at these distances, but increases substantially in field strength at intermediate and very short distances as a result of the larger tip-nanorod coupling. Due to the implementation of the non-local linear polarizability $\alpha_{xx}^r(x, x', \omega)$ for the nanorod, the spatial shapes of the two effective dipolar modes are preserved at all distances. In our model, vectorial dipole-dipole coupling hence results in near-fields, which are described as coherent, linear superpositions of the two initial, unperturbed modes. Note that in the present implementation our calculation does not account for the formation of localized gap modes, likely to be relevant for tip-sample distances of less than 3 nm.[5] For this, higher order multipole modes of tip and nanorod should be included in the model.

Figure S3b shows a series of normalized PNS spectra corresponding to the coupled near-fields in panel (a) with the detector placed under an out-of-plane angle of $\theta_d \simeq 10°$ (white arrow in a). Tip-sample distances are indicated below each curve. These data represent the modeling results that correspond to the experimental spectra presented in Figure 3d of our manuscript. By comparing the two sets of data, we find that our model quantitatively reproduces all major experimental features: At large and intermediate tip-sample distances beyond 20 nm we observe dispersive spectra. For short distances a clear transition to absorptive spectra is found. We also observe a clear spectral redshift and a local coupling-induced line broadening. As detailed in the manuscript, these effects can be attributed to different coupling mechanisms between tip and sample. While the line-broadening is a consequence of the **transverse** coupling between the nanorod and the axial resonance of the tip, mediated by the Green's tensor component $G_{xz}$, the spectral shifts orginate from the **longitudinal** coupling between the nanorod and the transverse blue-shifted resonances of the tip, mediated by $G_{xx}$.

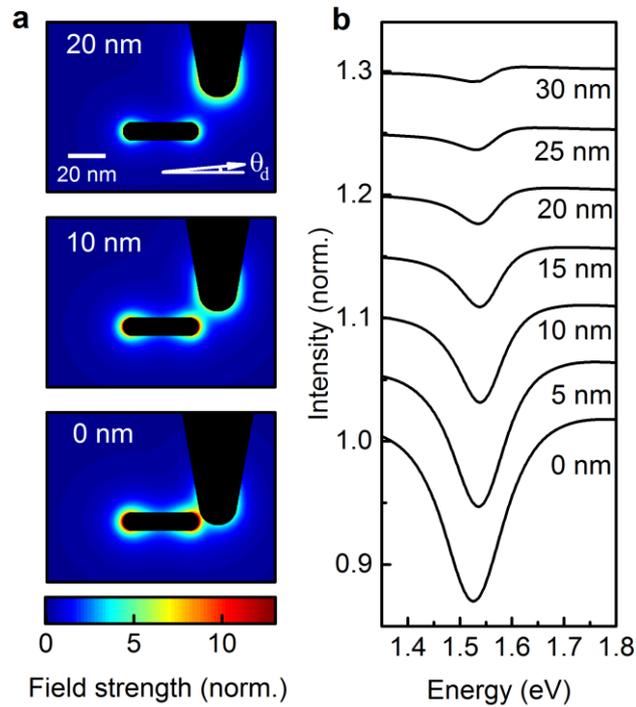

*Figure S3 Coupled dipole simulation of near-fields and PNS spectra along an approach curve.* **a,** *Near-fields around the SNOM taper and an individual gold nanorod for three different tip-sample distances (indicated at top left of each panel) along an approach curve. Only the tip is excited by a nanofocused field polarized along the tip axis with a field strength of* $1\,V/m$ *and at a photon energy of* $1.55\,eV$. *All model parameters are chosen as described in the main text.* **b,** *Corresponding simulated PNS spectra along the approach curve shown in (a). The detector position is indicated by the arrow in (a), tip-sample distances are indicated below each curve. Upon approaching tip and sample, a transition from dispersive to absorptive spectra, a red-shift in resonance and a line broadening are observed, all in quantitative agreement with the experimental spectra shown in Figure 3d of the main text. Subsequent spectra are vertically offset for clarity.*

To substantiate this assignment further, we repeated the simulations shown in Figure 5d-g of the main text, but with the polarizability tensor of the tip $\overleftrightarrow{\alpha^t}$ reduced to $\overleftrightarrow{\alpha_{zz}^t}$. We thus assume a scalar tip polarizability, restricted to the dominant longitudinal component of the tip polarizability. The transverse blue-shifted mode of the tip $\alpha_{xx/yy}^t(\omega)$ is completely switched off. As in the calculations in the main text, we simulated normalized PNS spectra as a function of position while scanning the tip across a single gold nanorod. By fitting a Fano-line following Equation (2) to each spectrum, we obtain two-dimensional parameter maps. These results are presented in Figure S4.

We observe that the scattering amplitude $|\mu|^2\gamma$ **(a)**, the Fano-phase $\phi$ **(b)** and the local linewidth $\gamma$ **(c)** show virtually the same spatial distribution and magnitude as in the simulations presented in Figure 5 of the main text. The map of the local resonance energy $\omega_0$ **(d)**, however, looks markedly different. While a small, local shift in resonance to lower energies on the left part of the nanorod and to higher energies on the right part persists, the dominant crescent-moon shaped red-shift at both apices of the rod that is seen in Figure 5g of the main text is completely absent. This shows that a scalar tip-coupling model, considering only the coupling between the longitudinal polarizability of the tip and the nanorod, cannot account for the experimental observations. Instead, the crescent moon shaped red-shift can only be reproduced if we assume that the longitudinal polarizability of the nanorod also couples to the transverse polarizability of the tip: the signature of vectorial near field coupling.

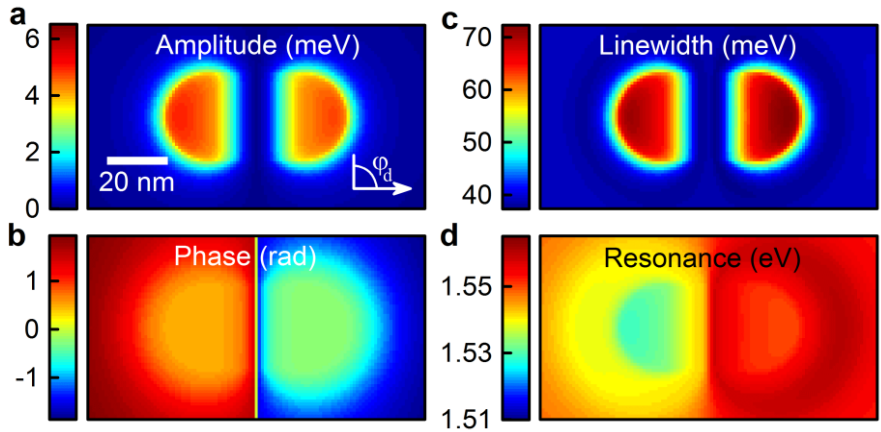

*Figure S4 Fano-line analysis of simulated PNS spectra using a scalar tip polarizability $\overleftrightarrow{\alpha^t_{zz}}$. a-d, Maps of the oscillator strength $|\mu|^2\gamma$ (a), Fano phase $\phi$ (b), linewidth $\gamma$ (c) and resonance energy $\omega_0$, (d) extracted from the coupled dipole model. All parameters are kept the same as for Figure 5(d-g) of the main text, only the weak, off-resonant transversal tip polarizability $\alpha^t_{xx} = \alpha^t_{yy}$ is set to zero. The local increase in linewidth in (c) caused by the coupling of the nanorod to the longitudinal dipolar polarization of the tip is still clearly observed, however, the strong red shifts near the rim of the rod (Figure 5g,h, main text) induced by the nanorod coupling to the transverse tip polarizability have completely disappeared.*

### 4. Perturbation series expansion of vectorial near-field coupling

In our manuscript we show that most of the dominant features observed in experimental PNS spectra are well understood by expanding the tip-nanorod interaction in a multiple scattering series: Dispersive line shapes are observed in first order scattering. When including second order scattering, a transition to absorptive spectra occurs. Spectral line broadening and local red shifts are obtained in higher orders of the perturbation series. Figure S5 schematically illustrates the dipole moments and field propagators involved in the first four orders of tip-sample scattering:

In zeroth order **(a)** the nanofocused SPP field $\mathbf{E}_0$ polarized along the nanotaper axis induces a dipole moment $\mathbf{p}_t^{(0)}(\omega) = \overleftrightarrow{\alpha^t}(\omega)\mathbf{E}_0$ at the position of the tip $\mathbf{r}_t$. This dipole then generates near- and far-fields $\mathbf{E}_t^{(0)}(\mathbf{r},\omega) = \frac{\omega^2}{\epsilon_0 c^2}\overleftrightarrow{\mathbf{G}}(\mathbf{r},\mathbf{r}_t,\omega)\mathbf{p}_t^{(0)}(\omega)$ with $\overleftrightarrow{\mathbf{G}}(\mathbf{r},\mathbf{r}_t,\omega)$ being the Green's propagator of free space. The far-field components impinge on the detector placed at $\mathbf{r} = \mathbf{r}_d$ resulting in a detected PNS spectrum $\left|\mathbf{E}_t^{(0)}(\mathbf{r}_d,\omega)\right|^2$ which reflects the unpertubed polarizability of the tip apex. In our model, this spectrum, together with the weak background field $\mathbf{E}_b(\omega,\mathbf{r}_d)$, plays the role of the experimental scattering spectrum on glass presented in Figure 1b and is used for normalization.

In first order **(b)** the zeroth order field $\mathbf{E}_t^{(0)}$ emitted by the tip propagates to the nanorod and its $x$-component induces a polarization along the long nanorod axis. With the discretized polarizability $\overleftrightarrow{\alpha^r}$ shown in Figure S1a, the resulting chain of $N$ dipole moments $\mathbf{p}_{r,i}^{(1)}$ at positions $x_i$ along the nanorod axis is given by $\mathbf{p}_{r,i}^{(1)}(\omega, x_i) = \sum_{j=1}^{N} \overleftrightarrow{\alpha^r}(x_i, x_j, \omega)\mathbf{E}_t^{(0)}(\mathbf{r}_j,\omega)$. Here, $\mathbf{r}_i = (x_i, 0, 0)$, i.e., the nanorod is centered at the coordinate origin. The chain of induced dipoles then re-radiates fields $\mathbf{E}_r^{(1)}(\mathbf{r},\omega) = \frac{\omega^2}{\epsilon_0 c^2}\sum_{i=1}^{N}\overleftrightarrow{\mathbf{G}}(\mathbf{r},\mathbf{r}_{x_i},\omega)\mathbf{p}_{r,i}^{(1)}(\omega, x_i)$. As discussed in the manuscript, close to the nanorod resonance the fields $\mathbf{E}_t^{(0)}$ and $\mathbf{E}_r^{(1)}$ are essentially $\pm 90°$ out of phase. The interference between these two fields thus results in the dispersive line shapes observed at larger tip-sample distances beyond 20 nm both in experiment and simulation (Figure 3d and S3b).

In second order **(c)** the fields $\mathbf{E}_r^{(1)}$ propagates back to the tip and induces a dipole moment $\mathbf{p}_t^{(2)}(\omega) = \overleftrightarrow{\alpha^t}(\omega)\mathbf{E}_r^{(1)}(\mathbf{r}_t,\omega)$ Since the second order field $\mathbf{E}_t^{(2)}(\mathbf{r},\omega) = \frac{\omega^2}{\epsilon_0 c^2}\overleftrightarrow{\mathbf{G}}(\mathbf{r},\mathbf{r}_t,\omega)\mathbf{p}_t^{(2)}(\omega)$ is phase shifted by another $\pm 90°$ with respect to $\mathbf{E}_r^{(1)}$, it destructively interferes with $\mathbf{E}_t^{(0)}$, giving rise to the transition from dispersive to absorptive line shapes observed at intermediate and short tip-sample distances. In the second order of the perturbation series, also the blue-detuned transversal resonance of the tip $\alpha_{xx/yy}^t(\omega)$ comes into play, since the nanorod field $\mathbf{E}_r^{(1)}$ contains components both in $z$- and $xy$-direction.

Panel **(d)** illustrates the third order scattering process, in which the field $\mathbf{E}_t^{(2)}$ induces dipole moments $\mathbf{p}_{r,i}^{(3)}(\omega, x_i)$ in the nanorod and the nanorod re-radiates fields $\mathbf{E}_r^{(3)}$. In this process, both transversal and longitudinal tip-sample coupling are involved. That is, the $x$-component of the tip-field contains contributions both from the axial and the transversal polarization induced into the tip in second order.

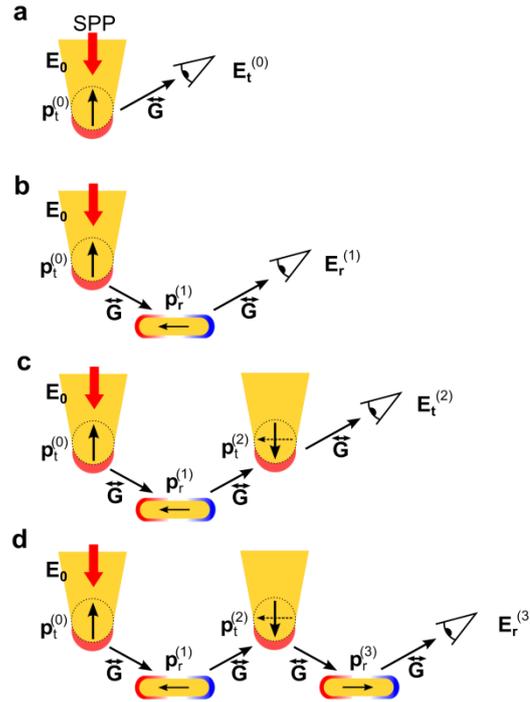

*Figure S5 Perturbative expansion of vectorial dipole-dipole coupling. a-d,* Schematic representation of zeroth to third-order scattering processes between two effective dipoles representing tip apex and nanorod. In zeroth order the nanofocused SPP field $\mathbf{E}_0$ induces a dipole moment $\mathbf{p}_t^{(0)}$ along the tip axis, which generates a field $\mathbf{E}_t^{(0)} \propto \overleftrightarrow{G}\overleftrightarrow{\alpha^t}\mathbf{E}_0$. In first order $\mathbf{E}_t^{(0)}$ then induces a dipole moment $\mathbf{p}_r^{(1)}$ along the nanorod which generates a secondary field $\mathbf{E}_r^{(1)} \propto \overleftrightarrow{G}\overleftrightarrow{\alpha^r}\overleftrightarrow{G}\overleftrightarrow{\alpha^t}\mathbf{E}_0$. In second order $\mathbf{E}_r^{(1)}$ acts back on the tip and induces a dipole $\mathbf{p}_t^{(2)}$ with components both along and across the taper, mediated by the vectorial coupling between tip and sample. This results in a field $\mathbf{E}_t^{(2)} \propto \overleftrightarrow{G}\overleftrightarrow{\alpha^t}\overleftrightarrow{G}\overleftrightarrow{\alpha^r}\overleftrightarrow{G}\overleftrightarrow{\alpha^t}\mathbf{E}_0$. The interference between $\mathbf{E}_t^{(0)}$ and $\mathbf{E}_r^{(1)}$ leads to dispersive lineshapes in far field detection. The destructive interference between $\mathbf{E}_t^{(0)}$ and $\mathbf{E}_t^{(2)}$ accounts for the observed transition to absorptive lines for larger coupling. For tip-sample distances below $10\ nm$ at least ten scattering orders are necessary to account for the experimentally observed coupling-induced local red-shift and increase in linewidth of the scattered fields.